\documentclass[floatfix,aps,prd,twocolumn,letterpaper,lengthcheck,superscriptaddress,showpacs,amssymb,amsmath,amsfonts,footinbib,altaffilletter,nopreprintnumbers,showpacs]{revtex4-1}
\pdfoutput=1

\usepackage{acronym}
\usepackage{booktabs}
\usepackage{amsmath,amssymb,graphicx}
\usepackage{color}
\usepackage[colorlinks, linkcolor=blue, pdfborder={0 0 0}, breaklinks=true]{hyperref}
\usepackage[utf8]{inputenc}
\usepackage[table]{xcolor}
\usepackage{tabularx}
\usepackage{siunitx}
\usepackage{subfigure}
\usepackage{xspace}

\DeclareSIUnit\parsec{pc}
\usepackage{scrextend}
\usepackage{textgreek}

\usepackage{graphicx}

\usepackage{makerobust}
\makeatletter
\@ifundefined{caption@xref}{}{%
  \MakeRobustCommand\caption@xref
}
\makeatother

\def\mnras{\ref@jnl{MNRAS}}             

\renewcommand{\today}{\number\day\space\ifcase\month\or
  January\or February\or March\or April\or May\or June\or
  July\or August\or September\or October\or November\or December\fi
  \space\number\year}

\begin{document}
\title{Automating the Inclusion of Subthreshold Signal-to-Noise Ratios for Rapid Gravitational-Wave Localization}

\author{Cody Messick}
\affiliation{The Center for Gravitational Physics, University of Texas, Austin, TX 78712, USA}

\author{Surabhi Sachdev}
\affiliation{Department of Physics, The Pennsylvania State University, University Park, PA 16802, USA}
\affiliation{Institute for Gravitation and the Cosmos, The Pennsylvania State University, University Park, PA 16802, USA}

\author{Kipp Cannon}
\affiliation{RESCEU, University of Tokyo, Tokyo, 113-0033, Japan}

\author{Sarah Caudill}
\affiliation{Nikhef, Science Park 105, 1098 XG Amsterdam, Netherlands}

\author{Chiwai Chan}
\affiliation{RESCEU, University of Tokyo, Tokyo, 113-0033, Japan}

\author{Jolien D. E. Creighton}
\affiliation{Leonard E.\ Parker Center for Gravitation, Cosmology, and Astrophysics, University of Wisconsin-Milwaukee, Milwaukee, WI 53201, USA}

\author{Ryan Everett}
\affiliation{Department of Physics, The Pennsylvania State University, University Park, PA 16802, USA}
\affiliation{Institute for Gravitation and the Cosmos, The Pennsylvania State University, University Park, PA 16802, USA}

\author{Becca Ewing} 
\affiliation{Department of Physics, The Pennsylvania State University, University Park, PA 16802, USA}
\affiliation{Institute for Gravitation and the Cosmos, The Pennsylvania State University, University Park, PA 16802, USA}

\author{Heather Fong}
\affiliation{RESCEU, University of Tokyo, Tokyo, 113-0033, Japan}

\author{Patrick Godwin}
\affiliation{Department of Physics, The Pennsylvania State University, University Park, PA 16802, USA}
\affiliation{Institute for Gravitation and the Cosmos, The Pennsylvania State University, University Park, PA 16802, USA}

\author{Chad Hanna}
\affiliation{Department of Physics, The Pennsylvania State University, University Park, PA 16802, USA}
\affiliation{Department of Astronomy \& Astrophysics, Pennsylvania State University, University Park, PA, 16802, USA}
\affiliation{Institute for Gravitation and the Cosmos, The Pennsylvania State University, University Park, PA 16802, USA}

\author{Rachael Huxford}
\affiliation{Department of Physics, The Pennsylvania State University, University Park, PA 16802, USA}
\affiliation{Institute for Gravitation and the Cosmos, The Pennsylvania State University, University Park, PA 16802, USA}

\author{Shasvath Kapadia}
\affiliation{Leonard E.\ Parker Center for Gravitation, Cosmology, and Astrophysics, University of Wisconsin-Milwaukee, Milwaukee, WI 53201, USA}

\author{Alvin K. Y. Li}
\affiliation{LIGO Laboratory, California Institute of Technology, MS 100-36, Pasadena, California 91125, USA}

\author{Rico K. L. Lo}
\affiliation{Department of Physics, The Chinese University of Hong Kong, Shatin, New Territories, Hong Kong}
\affiliation{LIGO Laboratory, California Institute of Technology, MS 100-36, Pasadena, California 91125, USA}

\author{Ryan Magee}
\affiliation{Department of Physics, The Pennsylvania State University, University Park, PA 16802, USA}
\affiliation{Institute for Gravitation and the Cosmos, The Pennsylvania State University, University Park, PA 16802, USA}

\author{Duncan Meacher}
\affiliation{Leonard E.\ Parker Center for Gravitation, Cosmology, and Astrophysics, University of Wisconsin-Milwaukee, Milwaukee, WI 53201, USA}

\author{Siddharth R. Mohite} 
\affiliation{Leonard E.\ Parker Center for Gravitation, Cosmology, and Astrophysics, University of Wisconsin-Milwaukee, Milwaukee, WI 53201, USA}

\author{Debnandini Mukherjee}
\affiliation{Department of Physics, The Pennsylvania State University, University Park, PA 16802, USA}
\affiliation{Institute for Gravitation and the Cosmos, The Pennsylvania State University, University Park, PA 16802, USA}

\author{Atsushi Nishizawa}
\affiliation{RESCEU, University of Tokyo, Tokyo, 113-0033, Japan}

\author{Hiroaki Ohta}
\affiliation{RESCEU, University of Tokyo, Tokyo, 113-0033, Japan}

\author{Alexander Pace}
\affiliation{Department of Physics, The Pennsylvania State University, University Park, PA 16802, USA}
\affiliation{Institute for Gravitation and the Cosmos, The Pennsylvania State University, University Park, PA 16802, USA}

\author{Amit Reza}
\affiliation{Department of Physics, Indian Institute of Technology Gandhinagar, Gujarat 382355, India}

\author{Minori Shikauchi} 
\affiliation{RESCEU, University of Tokyo, Tokyo, 113-0033, Japan}

\author{Leo Singer}
\affiliation{NASA/Goddard Space Flight Center, Greenbelt, MD 20771, USA}

\author{Divya Singh}
\affiliation{Department of Physics, The Pennsylvania State University, University Park, PA 16802, USA}
\affiliation{Institute for Gravitation and the Cosmos, The Pennsylvania State University, University Park, PA 16802, USA}

\author{Javed Rana SK} 
\affiliation{Department of Physics, The Pennsylvania State University, University Park, PA 16802, USA}
\affiliation{Institute for Gravitation and the Cosmos, The Pennsylvania State University, University Park, PA 16802, USA}

\author{Leo Tsukada}
\affiliation{RESCEU, University of Tokyo, Tokyo, 113-0033, Japan}

\author{Daichi Tsuna}
\affiliation{RESCEU, University of Tokyo, Tokyo, 113-0033, Japan}

\author{Takuya Tsutsui}
\affiliation{RESCEU, University of Tokyo, Tokyo, 113-0033, Japan}

\author{Koh Ueno}
\affiliation{RESCEU, University of Tokyo, Tokyo, 113-0033, Japan}

\author{Aaron Zimmerman}
\affiliation{The Center for Gravitational Physics, University of Texas, Austin, TX 78712, USA}

%\date[\relax]{Compiled: \today}

\begin{abstract}
The accurate localization of gravitational-wave (GW) events in low-latency is a
crucial element in the search for further multimessenger signals from these
cataclysmic events. The localization of these events in low-latency uses
signal-to-noise ratio (SNR) time-series from matched-filtered searches which
identify candidate events.  Here we report on an improvement to the
GstLAL-based inspiral pipeline, the low-latency pipeline that identified
GW170817 and GW190425, which automates the use of SNRs from all detectors in
the network in rapid localization of GW events.  This improvement was
incorporated into the detection pipeline prior to the recent third observing
run of the Advanced LIGO and Advanced Virgo detector network.  Previously for
this pipeline, manual intervention was required to use SNRs from all detectors
if a candidate GW event was below an SNR threshold for any detector in the
network.  The use of SNRs from subthreshold events can meaningfully decrease
the area of the 90\% confidence region estimated by rapid localization.  To
demonstrate this, we present a study of the simulated detections of
$\mathcal{O}(2\times10^4)$ binary neutron stars using a network mirroring the
second observational run of the Advanced LIGO and Virgo detectors.  When
incorporating subthreshold SNRs in rapid localization, we find that the
fraction of events that can be localized down to $100~\si{\deg^2}$ or smaller
increases by a factor 1.18.
\end{abstract}

% NOTE I just kept the pacs from the catalog paper, may want to revisit

\maketitle

% ======================
%  ACRONYMS
% ======================
\acrodef{LSC}[LSC]{LIGO Scientific Collaboration}
\acrodef{aLIGO}{Advanced Laser Interferometer Gravitational wave Observatory}
\acrodef{aVirgo}{Advanced Virgo}
\acrodef{LIGO}[LIGO]{Laser Interferometer Gravitational-Wave Observatory}
\acrodef{IFO}[IFO]{interferometer}
\acrodef{LHO}[LHO]{LIGO-Hanford}
\acrodef{LLO}[LLO]{LIGO-Livingston}
\acrodef{O2}[O2]{second observing run}
\acrodef{O1}[O1]{first observing run}

\acrodef{BH}[BH]{black hole}
\acrodef{BBH}[BBH]{binary black hole}
\acrodef{BNS}[BNS]{binary neutron star}
\acrodef{NS}[NS]{neutron star}
\acrodef{BHNS}[BHNS]{black hole--neutron star binaries}
\acrodef{NSBH}[NSBH]{neutron star--black hole binary}
\acrodef{PBH}[PBH]{primordial black hole binaries}
\acrodef{CBC}[CBC]{compact binary coalescence}
\acrodef{GW}[GW]{gravitational wave}

\acrodef{SNR}[SNR]{signal-to-noise ratio}
\acrodef{FAR}[FAR]{false alarm rate}
\acrodef{IFAR}[IFAR]{inverse false alarm rate}
\acrodef{FAP}[FAP]{false alarm probability}
\acrodef{PSD}[PSD]{power spectral density}

\acrodef{GR}[GR]{general relativity}
\acrodef{NR}[NR]{numerical relativity}
\acrodef{PN}[PN]{post-Newtonian}
\acrodef{EOB}[EOB]{effective-one-body}
\acrodef{ROM}[ROM]{reduced-order-model}
\acrodef{IMR}[IMR]{inspiral-merger-ringdown}

\acrodef{PDF}[PDF]{probability density function}
\acrodef{PE}[PE]{parameter estimation}
\acrodef{CL}[CL]{credible level}

\acrodef{LAL}[LAL]{LIGO Algorithm Library}

\newcommand{\PN}[0]{\ac{PN}\xspace}
\newcommand{\BBH}[0]{\ac{BBH}\xspace}
\newcommand{\BNS}[0]{\ac{BNS}\xspace}
\newcommand{\BH}[0]{\ac{BH}\xspace}
\newcommand{\NR}[0]{\ac{NR}\xspace}
\newcommand{\GRW}[0]{\ac{GW}\xspace}
\newcommand{\SNR}[0]{\ac{SNR}\xspace}
\newcommand{\aLIGO}[0]{\ac{aLIGO}\xspace}
\newcommand{\PE}[0]{\ac{PE}\xspace}
\newcommand{\IMR}[0]{\ac{IMR}\xspace}
\newcommand{\PDF}[0]{\ac{PDF}\xspace}
\newcommand{\GR}[0]{\ac{GR}\xspace}
\newcommand{\PSD}[0]{\ac{PSD}\xspace}
\newcommand{\EOS}[0]{\ac{EOS}\xspace}

\section{Introduction}
The GstLAL-based inspiral pipeline, referred to as GstLAL in this paper for
brevity, is a matched-filtering analysis pipeline designed to search for
\acp{GW} from compact binary
inspirals~\cite{sachdev2019gstlal,messick2017analysis,cannon2012toward}. It
operates in two modes, low or high latency, and is built using GStreamer, a
multimedia streaming framework~\cite{gstreamer}.  The low-latency GstLAL
analysis was the first to identify both GW170817~\cite{abbott2017gw170817} and
GW190425~\cite{abbott2020gw190425}, the only two \ac{BNS} mergers confidently
identified via \acp{GW} to date; it was also the first to identify
GW151226~\cite{abbott2016gw151226}, the second confident \ac{BBH} merger ever
detected. In addition to GstLAL, there are three other low-latency
matched-filtering piplines searching for \acp{GW} from compact binary
inspirals: PyCBC Live~\cite{nitz2018rapid}, MBTA~\cite{adams2016low}, and
SPIIR~\cite{hooper2012summed,chu2017low}.

One of the primary motivations to search for \acp{GW} in low-latency is to
enable searches for counterparts by the global astronomy community. Beginning
in Advanced LIGO's~\cite{aasi2015advanced} third and Advanced
Virgo's~\cite{acernese2014advanced} second observing run, O3, any low-latency
compact binary merger observation that passed a \ac{FAR} threshold of 1 per 10
months generated a public alert~\cite{LIGOEMFOLLOWUSERGUIDE}. The preliminary
notice, the first public alert generated after the
merger~\cite{LIGOEMFOLLOWUSERGUIDE}, for candidates identified by compact
binary matched-filter pipelines typically contained a localization estimate
produced by BAYESTAR~\cite{singer2016rapid}. BAYESTAR estimates the position of
the source using a short duration of matched-filter \SNR time-series
surrounding the time of the candidate from each
detector~\cite{singer2016rapid}. Information about the candidate is uploaded to
the Gravitational-Wave Candidate Event Database (GraceDB)~\cite{gracedb}, which
triggers the BAYESTAR localization~\cite{LIGOEMFOLLOWUSERGUIDE}.

\begin{figure}
\subfigure{\includegraphics[width=\linewidth]{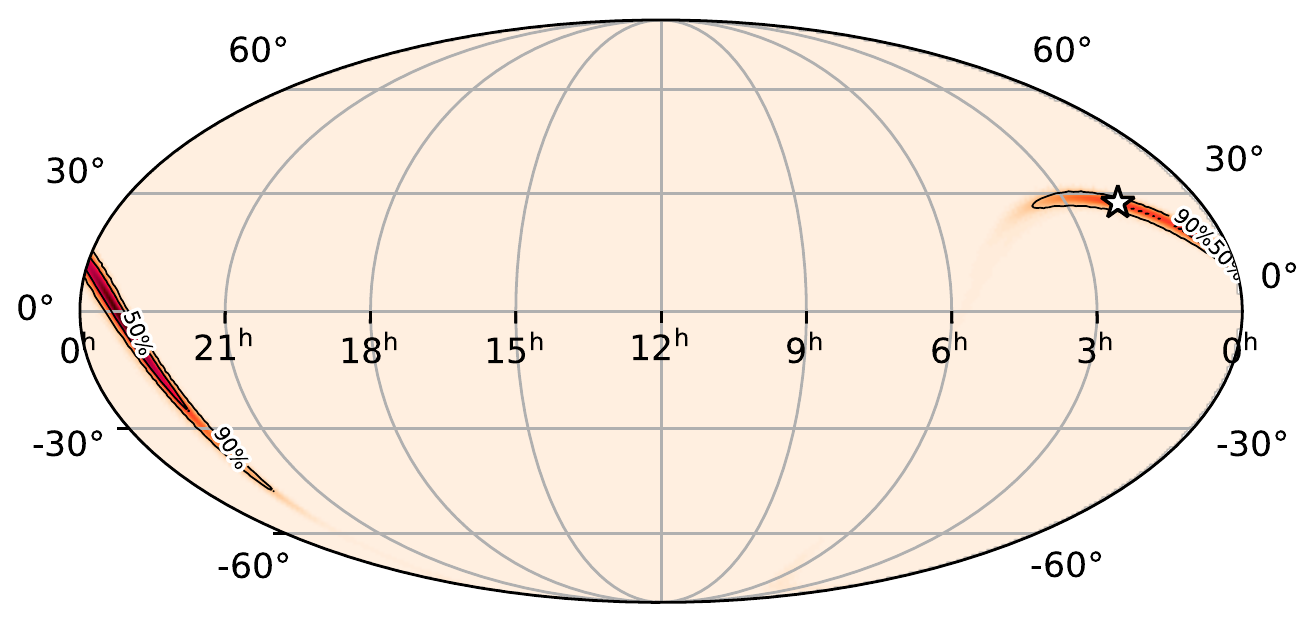}}
\subfigure{\includegraphics[width=\linewidth]{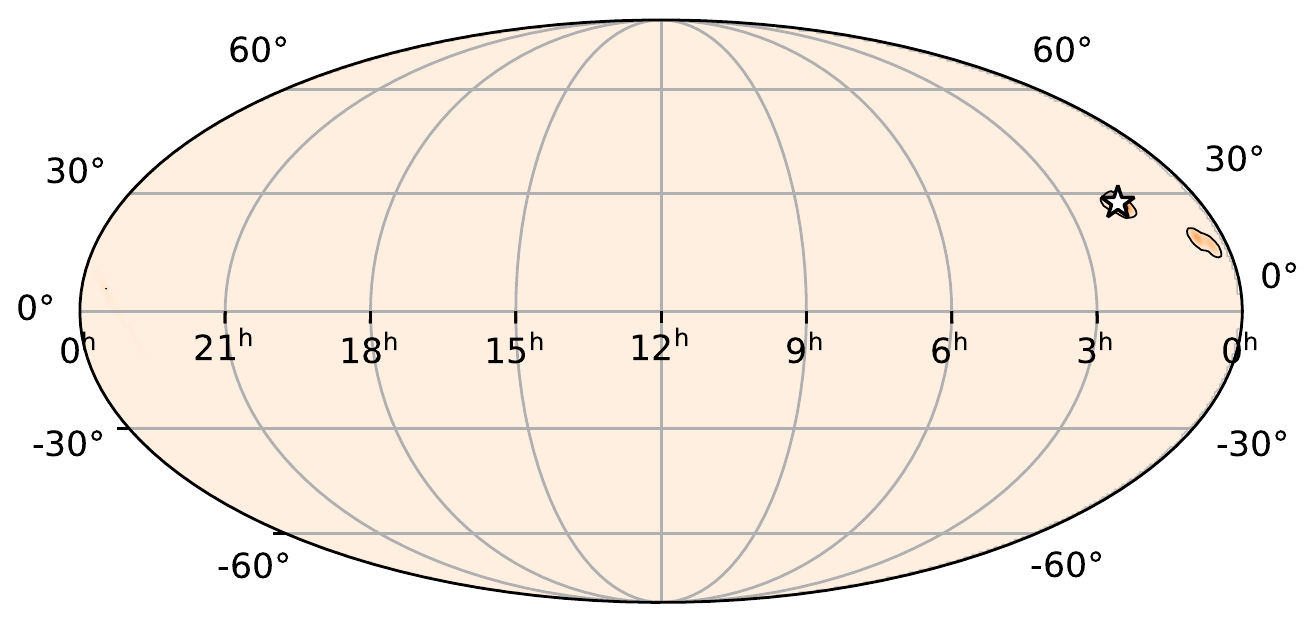}}
\subfigure{\includegraphics[width=\linewidth]{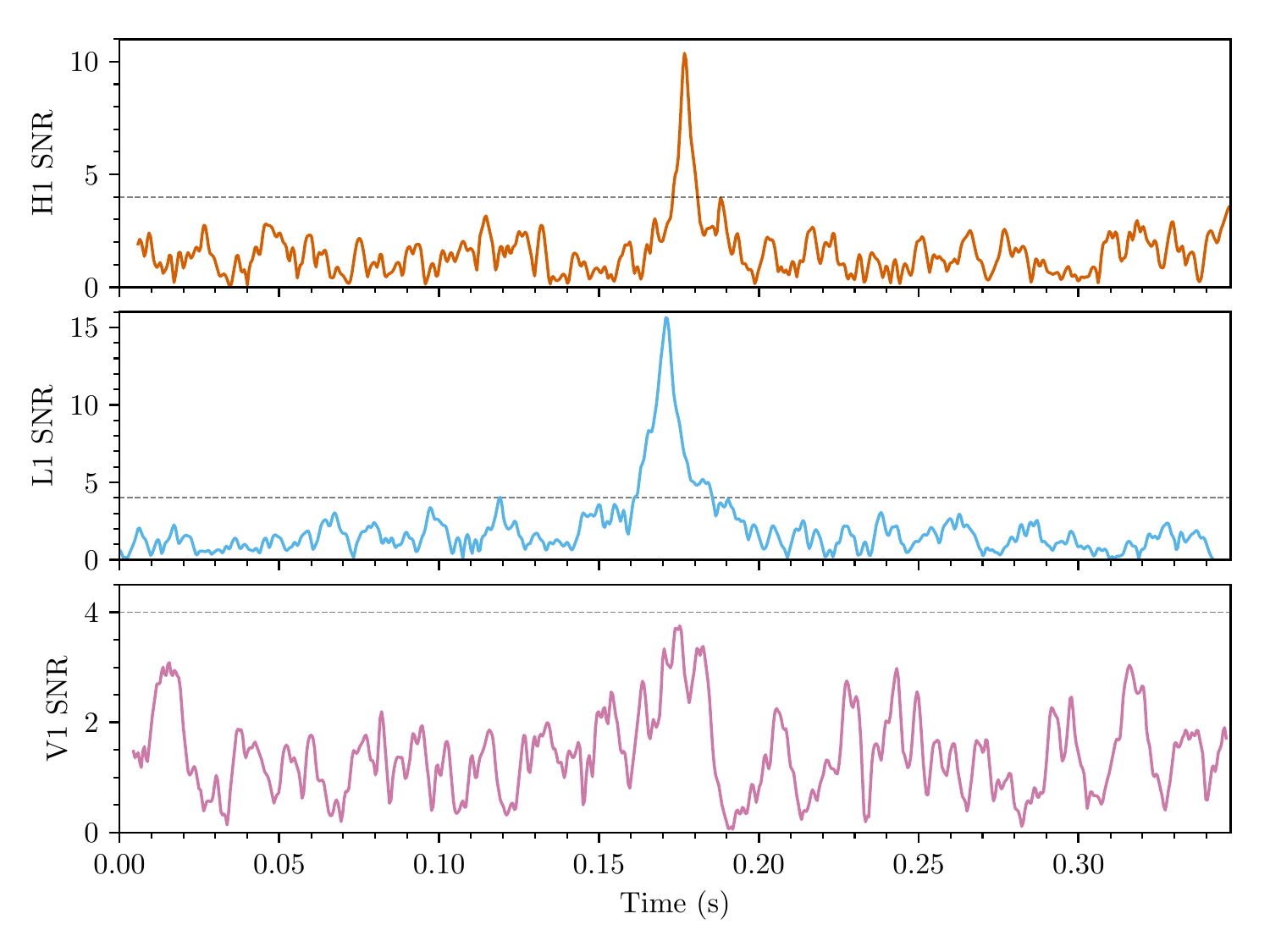}}
\caption{Top: Localization of a simulated \ac{BNS} merger estimated by
BAYESTAR using only \ac{SNR} data from detectors that produced \ac{SNR} peaks
above threshold. The star indicates the source of the simulated signal. The
total area of the 90\% confidence region is 527~$\si{\deg^2}$. Middle: The
localization of the same simulated signal, but estimated using \ac{SNR} from
all available detectors.  The total area of the 90\% confidence region is
85~$\si{\deg^2}$.  Bottom: The \ac{SNR} time-series in each detector made
available to use for localization. The dashed line shows the threshold. The
network SNR without Virgo is 18.8, with Virgo it is 19.2}
\label{fig:skymap}
\end{figure}

Although the localization performed by BAYESTAR benefits from having all
available \ac{SNR}, GstLAL only considers \acp{SNR} above a threshold $\rho^*$
when estimating the significance of a candidate. Imposing an \ac{SNR} threshold
drastically reduces the number of candidates to consider, hence reducing the
computational cost of performing the analysis, while only marginally reducing
the sensitivity of the analysis to signals. Prior to O3, acquiring subthreshold
\acp{SNR} from GstLAL required manual intervention, meaning the low-latency
analysis could not automatically upload all of the \ac{SNR} available in cases
where the maximum \ac{SNR} in a given detector is below threshold. This was a
small effect until the end of O2, when Advanced Virgo joined the observing run
with a lower sensitivity than the two advanced LIGO
detectors~\cite{ligo2018gwtc}. GW170817 was detected shortly after Virgo joined
O2, and although the \ac{SNR} in Virgo was subthreshold, { \it i.e.\ }below
$\rho^*$, the addition of Virgo decreased the 90\% confidence region estimated
by BAYESTAR from 190~$\si{\deg^2}$ to
31~$\si{\deg^2}$~\cite{abbott2017gw170817}. Cases like this will become more
common as more ground-based detectors of varying sensitivities join the
worldwide network of detectors~\cite{Aasi:2013wya}, {\it e.g.\
}KAGRA~\cite{akutsu2020overview}. An example of this effect for a simulated
\ac{BNS} signal in simulated Gaussian data can be seen in
Fig.~\ref{fig:skymap}, the top panel shows the localization estimated by
BAYESTAR when only using the advanced LIGO detectors, the middle panel shows
the localization when the \ac{SNR} from Virgo, which is below threshold, is
included, and the bottom panel shows the \ac{SNR} data used in the
localization.

GstLAL began providing subthreshold \acp{SNR} for rapid localization
automatically at the beginning of O3, although they are not used when
estimating the significance of candidates. The GStreamer element in the GstLAL
pipeline that applies the \ac{SNR} threshold, referred to as the trigger
generator, needed to be redesigned in order to automate the inclusion of
subthreshold \acp{SNR}.  This requirement is specific to GstLAL. PyCBC Live
began providing subthreshold \acp{SNR} automatically by the end of O2. In this
paper, we discuss the redesign of the trigger generator and why it was
necessary to automate the inclusion of subthreshold \acp{SNR} for rapid
localization. We then present the results of a study performed comparing rapid
localization of \ac{BNS} signals using BAYESTAR with and without subthreshold
\acp{SNR}.

\section{Automating the Inclusion of Subthreshold Signal-to-Noise Ratios}
GstLAL computes the \ac{SNR} time-series for each template, or model waveform,
and detector at a user-specified rate, typically \SI{2048}{\Hz}, as described
in
Refs.~\cite{sachdev2019gstlal,messick2017analysis,cannon2012toward,cannon2010singular}.
`Triggers', which consist of a timestamp, an \ac{SNR}, an autocorrelation-based
test-statistic $\xi^2$~\cite{sachdev2019gstlal,messick2017analysis}, and a
template ID, are produced by maximizing the \ac{SNR} over \SI{1}{\second}
windows for each template and computing $\xi^2$ using the surrounding \ac{SNR}
data from that template and
detector~\cite{sachdev2019gstlal,messick2017analysis,cannon2012toward,cannon2010singular}.
If the maximum \ac{SNR} is less than the user-specfied threshold $\rho^*$, a
trigger is not produced. Prior to the work presented here, the \ac{SNR} data
were discarded at this stage for templates that did not produce an \ac{SNR}
peak above threshold, while the \ac{SNR} data used to compute $\xi^2$,
typically $\mathcal{O}(100)$~\si{\ms}, were held in memory for templates that
did produce triggers in a given detector. The duration of \ac{SNR} saved is
notably longer than the maximum light-travel-time between detectors and the
duration of \ac{SNR} required by BAYESTAR for localization. Due to only saving
the \ac{SNR} surrounding a peak above threshold, the only \ac{SNR} data
available to BAYESTAR automatically were from detectors where the maximum
\ac{SNR} in that second was at least as large as $\rho^*$. In order to obtain
the \ac{SNR} from other detectors that were operating at the time but did not
produce an \ac{SNR} peak above the threshold, the \ac{SNR} generation stage
needed to be manually rerun and configured to write the \ac{SNR} time-series at
a user-provided time to disk before discarding from memory.

\begin{figure}
\subfigure{\includegraphics[width=\linewidth]{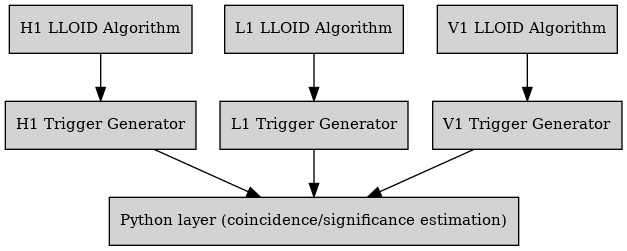}}
\subfigure{\includegraphics[width=\linewidth]{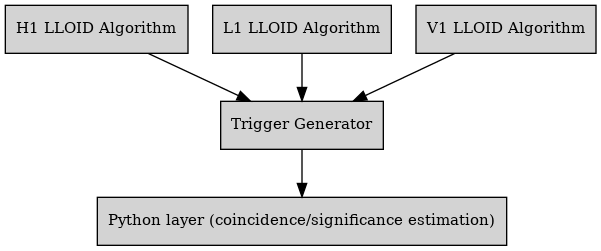}}
\caption{A view of the old (top) and O3 (bottom) workflows that generate
triggers in the GstLAL-based inspiral pipeline. H1, L1, and V1 refer to the
Hanford and Livingston LIGO observatories and the Virgo observatory
respectively. The LLOID Algorithm~\cite{cannon2012toward} computes an \ac{SNR}
time-series at a user-provided rate for each detector, which are then passed to
trigger generators that maximize the \ac{SNR} over 1~\si{\second} intervals and
form a trigger if the maximum passes a user-provided threshold, $\rho^*$
(typically 4). Top: Before O3, a different trigger generator was used for each
detector. A small interval of \ac{SNR} data centered around each trigger were
saved by the trigger generator, while the rest were discarded. As a
consequence, \ac{SNR} data from detectors that did not produce a peak above
threshold were not automatically available for rapid localization performed by
BAYESTAR~\cite{singer2016rapid}. Bottom: Beginning in O3, one trigger generator
ingests the \ac{SNR} data from all available detectors. A small interval of
\ac{SNR} data centered around each trigger are now saved by the trigger
generator for each detector, { \it i.e.\ } if a peak above threshold is
identified in one detector then \ac{SNR} data surrounding the time of that peak
are saved from all available detectors.  This ensures that rapid localization
performed by BAYESTAR in
low-latency~\cite{singer2016rapid,LIGOEMFOLLOWUSERGUIDE} automatically has
access to all available \ac{SNR} data.} \label{fig:workflow}
\end{figure}

The reason that the \ac{SNR} data were discarded by the trigger generator is
that it only considered \ac{SNR} data from one detector, { \it i.e.\ }each
detectors' \ac{SNR} data were ingested by a different instance of the trigger
generator and the instances had no knowledge of each other, as seen in the top
graphic in Fig.~\ref{fig:workflow}. This was an intentional design choice made
early in development to avoid the complexity of aggregating multiple
time-series streams. However, the GStreamer libraries have undergone many
changes since the original design and now include a stable class,
GstAggregator, that makes synchronizing multiple time-series easier.  The new
trigger generator is a subclass of GstAggregator, and now considers \ac{SNR}
data from all available detectors. If the maximum \ac{SNR} in a detector is
larger than the threshold $\rho^*$, \ac{SNR} data surrounding that peak are now
saved from all detectors, { \it e.g.\ }if an above-threshold peak \ac{SNR} is
identified in the Hanford LIGO detector, \acp{SNR} surrounding that time are
saved from all available detectors. The duration saved matches the interval
used to compute the trigger's $\xi^2$ test-statistic.  The new architecture of
the algorithm is depicted in the bottom panel of Fig.~\ref{fig:workflow}.

Ingesting multiple time-series required changing several aspects of the trigger
generator's algorithm. The internal algorithm that searches for \ac{SNR} peaks
above threshold is unchanged, and is still executed for individual detectors.
The primary differences between the original and new
trigger generators are: 1)~more metadata are saved, { \it e.g.\ }the metadata
required to retrieve \ac{SNR} from the other detectors when a peak is
identified above threshold; 2)~the selection of the interval over which to
search for \ac{SNR} peaks has changed, previously the intervals would bookend
gaps and could be different for each detector, but now the same interval is
used for all detectors and may include sub-second gaps; and 3)~the structure of
the buffers output by the element has changed because each output buffer now
contains more information than the original trigger generator's output buffers.
The last difference also required changes to the elements downstream of the
trigger generator in the GstLAL pipeline, { \it e.g.\ }new code was written to
extract subthreshold \ac{SNR} data that were not previously output by the
trigger generator.

\section{Injection Study}
Subthreshold \acp{SNR} can be vital in localizing \acp{GW} from compact binary
mergers. For example, the low \ac{SNR} of the Virgo trigger associated with
GW170817 reduced the area of the 90\% confidence region containing the source
as estimated by BAYESTAR from 190~$\si{\deg^2}$ to
31~$\si{\deg^2}$~\cite{abbott2017gw170817}. Although subthreshold triggers from
the GstLAL could be generated in previous observing runs, the manual
intervention required added minutes to the latency of producing a localization
estimate, a process that can be accomplished in seconds~\cite{singer2016rapid}. 

In order to quantify the effect that
subthreshold \acp{SNR} have on rapid localization by BAYESTAR, we inserted
$\mathcal{O}(2 \times 10^4)$ simulated \ac{BNS} signals into 13.0 \si{\day} of
simulated Gaussian data. The Gaussian data were re-colored using public noise
curves released by the LIGO Scientific and Virgo Collaborations that are
considered representative of the instruments' sensitivities during
O2~\cite{publicnoise}. The component masses of the simulated \ac{BNS} signals
are drawn from uniform distributions between $1 \; M_\odot$ and $3 \; M_\odot$,
the dimensionless spins are aligned with the binaries' orbital angular momenta
and drawn from uniform distributions from -0.05 to 0.05, the inclinations are
drawn from a uniform distribution, and the distances are drawn from a uniform
distribution from $20 \; \mathrm{Mpc}$ to $800 \; \mathrm{Mpc}$. These
distributions are not astrophysical, however the distance distribution is
expected to generate many signals below an \ac{SNR} threshold of $\rho^*=4$ in
Virgo given the network sensitivity in O2~\cite{ligo2018gwtc}.

\begin{figure}[t]
\includegraphics[width=\linewidth]{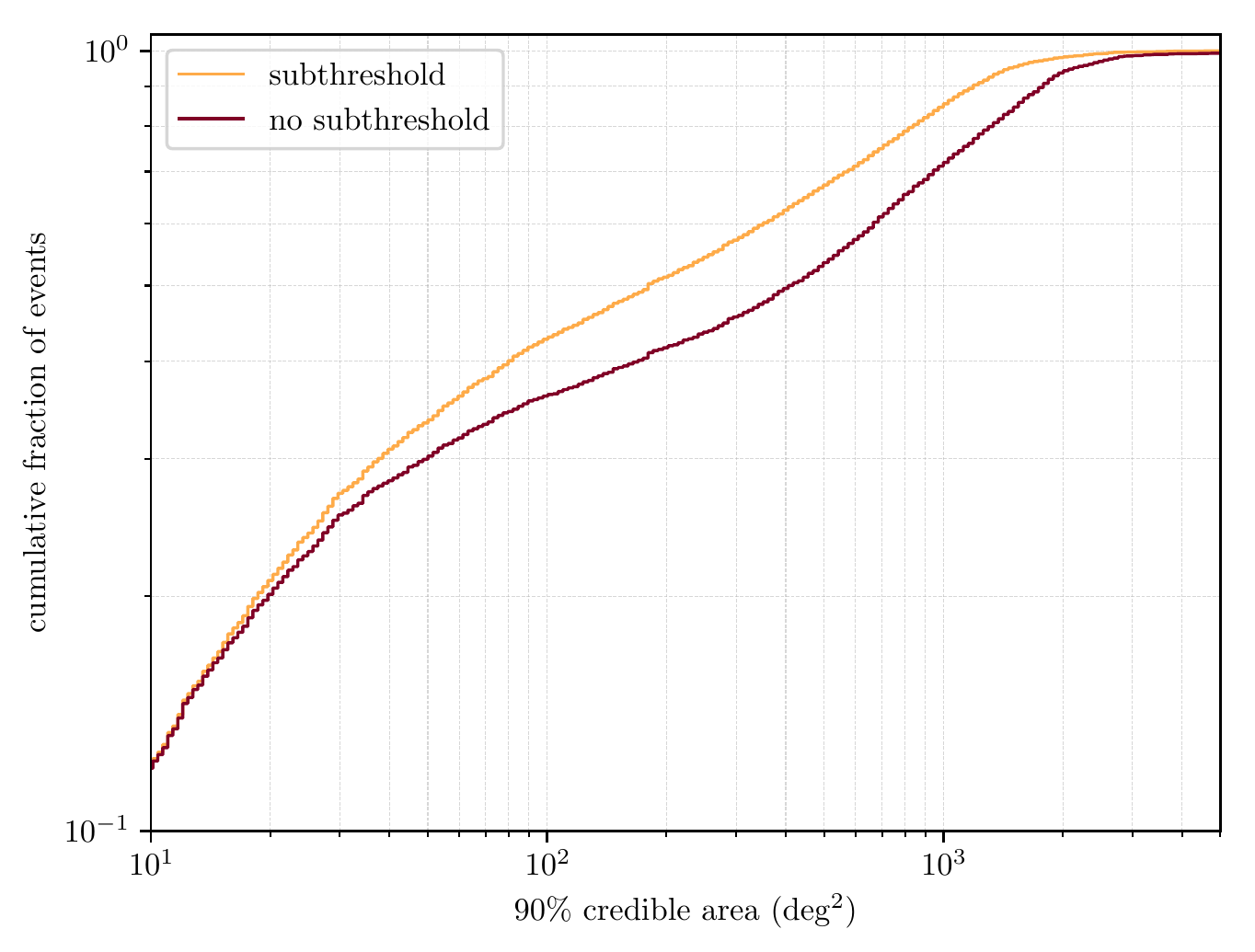}
\caption{The fractional cumulative number of events localized to a 90\%
credible region as a function of the 90\% credible region area. The events here
are $\mathcal{O}(2 \times 10^4)$ fake \ac{BNS} signals injected into 13.0 days
of Gaussian data that were re-colored using publicly available noise curves
that represent the sensitivity of the Advanced LIGO and Advanced Virgo
detectors during O2. The line labeled ``subthreshold'' shows the cumulative
number of events when including subthreshold \acp{SNR} in the localization
calculation performed by BAYESTAR~\cite{singer2016rapid}, while the line labled
``no subthreshold'' shows the cumulative number of events when subthreshold
\acp{SNR} are not used by BAYESTAR. Prior to the work presented here, inclusion
of subthreshold \acp{SNR} when localizing candidates identified by the
GstLAL-based inspiral pipeline required manual intervention which added
$\mathcal{O}(\mathrm{minutes})$ of latency to a process that can take only
seconds~\cite{singer2016rapid}. The cumulative fraction of events that are
localized to a region smaller than $100~\si{\deg^2}$ increased from $\sim 36\%$
to $\sim 43\%$ upon the inclusion of \acp{SNR}.}
\label{fig:moneyplot}
\end{figure}

\begin{figure}
\subfigure{\includegraphics[width=\linewidth]{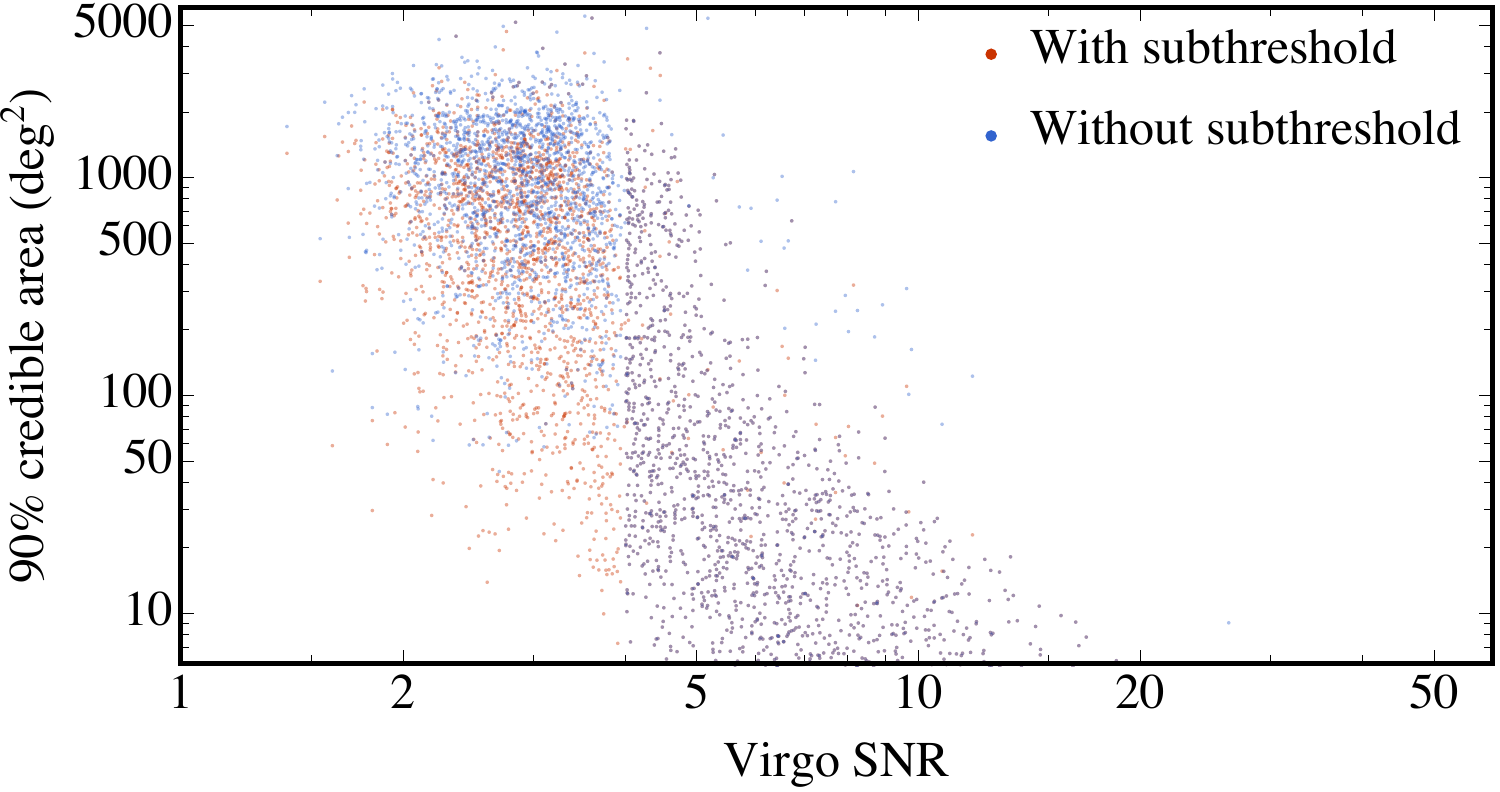}}
\subfigure{\includegraphics[width=\linewidth]{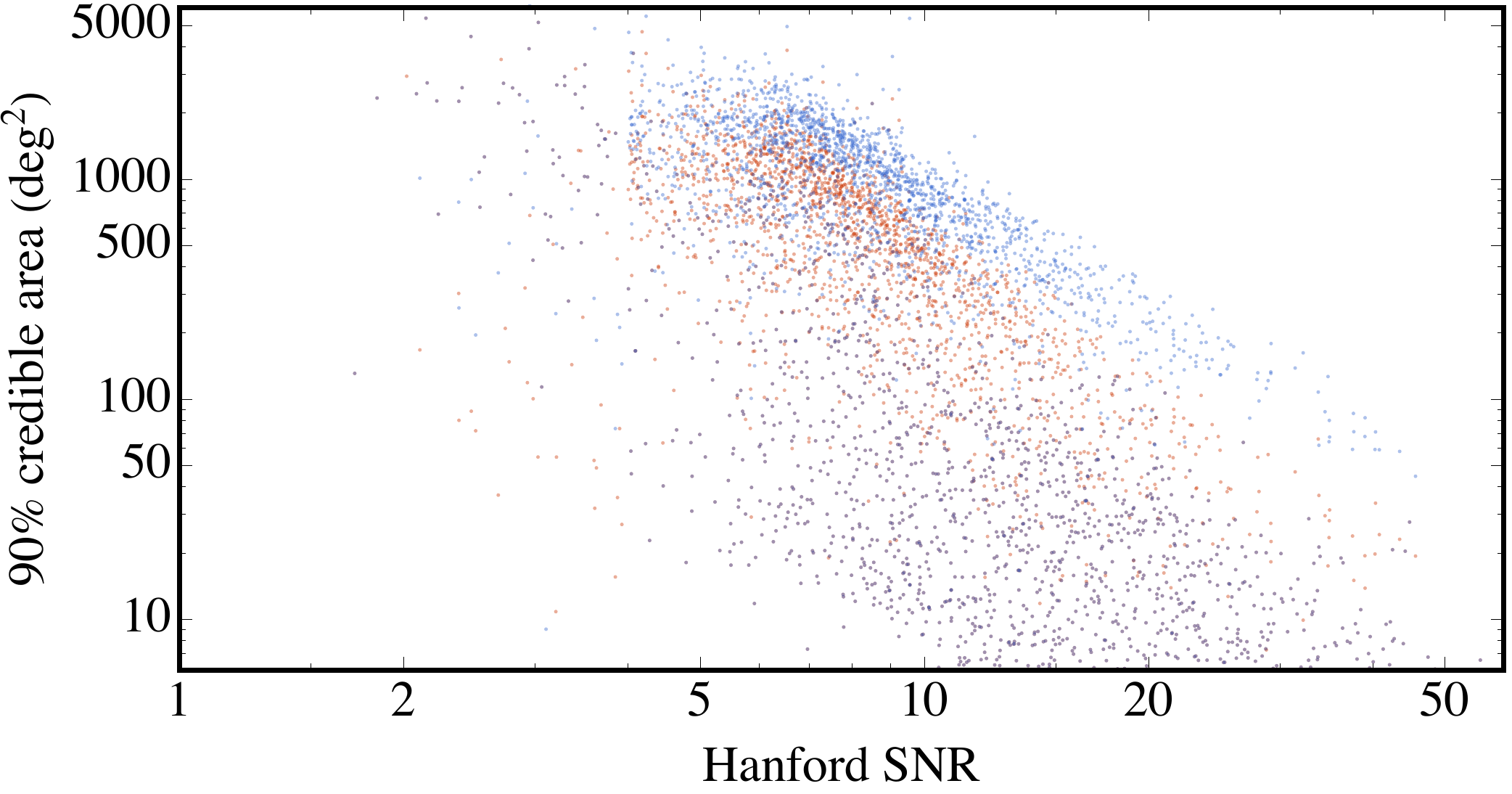}}
\caption{Scatter plots of the 90\% confidence areas estimated by BAYESTAR for
the recovered simulated \ac{BNS} signals. Top: The areas compared to the
\ac{SNR} in Virgo. The \ac{SNR} threshold used for this search was $\rho^*=4$.
A clear improvement in area is observed for Virgo triggers below threshold,
which is the expected behavior given the relative detector sensitivities.
Bottom: The areas compared to the \ac{SNR} in the Hanford, WA LIGO detector for
comparison to the top panel.}\label{fig:areavssnr}
\end{figure}

The data with simulated signals were analyzed using the version of GstLAL used
to produce the offline significance estimates in
Refs.~\cite{ligo2020gw190412,abbott2020gw190521,abbott2020gw190814}.  The
search was performed using the \ac{BNS} region of the template bank used in the
same searches~\cite{ligo2020gw190412,abbott2020gw190521,abbott2020gw190814},
with component masses in the interval {$[1,3) \; M_\odot$} and
aligned dimensionless-spin parameters in the interval
$[-0.05,0.05]$. As shown in Fig.~\ref{fig:moneyplot}, the
inclusion of subthreshold \acp{SNR} decreases the area of the 90\% confidence
regions for candidates that can be localized down to as low as $\sim
10~\si{\deg^2}$ or as high as $\sim 10^3~\si{\deg^2}$. This can be understood
qualitatively as the extremes on either end of the distribution, \textit{i.e.}
the signals that can be localized to better than $\sim 10~\si{\deg^2}$ or worse
than $\sim 10^3~\si{\deg^2}$, are made up by either extremely loud signals that
would not produce subthreshold \acp{SNR} or extremely quiet signals that are
barely observable. This can further be understood by comparing the area of the
90\% confidence regions with and without subthreshold \acp{SNR} against the
Virgo \ac{SNR}, as seen in the top panel of Fig.~\ref{fig:areavssnr}. The
majority of the decreased area occurs in cases with subthreshold Virgo
\acp{SNR}; similar behavior is not seen when comparing the sky areas to the
\ac{SNR} in one of the more sensitive LIGO detectors.  This is the result of a
network of detectors with inhomogeneous sensitivities, which is what we expect
for the foreseeable future.  Note that the visible gaps around an \ac{SNR} of 4
in Fig.~\ref{fig:areavssnr} are due to an artifact in how the peak \ac{SNR} is
computed for peaks below threshold. When a peak is above threshold, GstLAL
interpolates the \ac{SNR} time-series to improve the estimate of \ac{SNR},
which gives a slight increase in the \ac{SNR}. This interpolation is not done
for the subthreshold case, resulting in a gap just above the threshold
$\rho^*$, but the full \ac{SNR} time-series is used in the same way to localize
the event in both cases. The fraction of candidates localized to
$100~\si{\deg^2}$ or smaller increased from $\sim 36\%$ to $\sim 43\%$, an
increase of $\sim 18 \%$ relative to the rate without subthreshold \acp{SNR}. 

\section{Conclusion}
We presented work performed to automate the inclusion of subthreshold \acp{SNR}
into GraceDB~\cite{gracedb} uploads for use in localization by
BAYESTAR~\cite{singer2016rapid}. The original and current workflows that
generate triggers and dictate what \acp{SNR} are preserved in GstLAL are shown
in Fig.~\ref{fig:workflow}, and a study comparing the localization with and
without subthreshold \acp{SNR} is shown in Fig.~\ref{fig:moneyplot}. The rate
at which candidates were localized to $100~\si{\deg^2}$ or less increased from
$\sim 36\%$ to $\sim 43\%$. The new trigger generator described here was used
by GstLAL for the entirety of O3.

\acknowledgments
We gratefully acknowledge the support of the Eberly Research Funds of Penn
State and the National Science Foundation through OAC-1841480, PHY-1454389, and
PHY-1912578. The authors are grateful for computational resources provided by
the LIGO Lab and supported by the National Science Foundation through
PHY-1626190 and PHY-1700765. This document has LIGO document number: P2000445.

\bibliography{bibliography}

\begin{thebibliography}{24}%
\makeatletter
\providecommand \@ifxundefined [1]{%
 \@ifx{#1\undefined}
}%
\providecommand \@ifnum [1]{%
 \ifnum #1\expandafter \@firstoftwo
 \else \expandafter \@secondoftwo
 \fi
}%
\providecommand \@ifx [1]{%
 \ifx #1\expandafter \@firstoftwo
 \else \expandafter \@secondoftwo
 \fi
}%
\providecommand \natexlab [1]{#1}%
\providecommand \enquote  [1]{``#1''}%
\providecommand \bibnamefont  [1]{#1}%
\providecommand \bibfnamefont [1]{#1}%
\providecommand \citenamefont [1]{#1}%
\providecommand \href@noop [0]{\@secondoftwo}%
\providecommand \href [0]{\begingroup \@sanitize@url \@href}%
\providecommand \@href[1]{\@@startlink{#1}\@@href}%
\providecommand \@@href[1]{\endgroup#1\@@endlink}%
\providecommand \@sanitize@url [0]{\catcode `\\12\catcode `\$12\catcode
  `\&12\catcode `\#12\catcode `\^12\catcode `\_12\catcode `\%12\relax}%
\providecommand \@@startlink[1]{}%
\providecommand \@@endlink[0]{}%
\providecommand \url  [0]{\begingroup\@sanitize@url \@url }%
\providecommand \@url [1]{\endgroup\@href {#1}{\urlprefix }}%
\providecommand \urlprefix  [0]{URL }%
\providecommand \Eprint [0]{\href }%
\providecommand \doibase [0]{http://dx.doi.org/}%
\providecommand \selectlanguage [0]{\@gobble}%
\providecommand \bibinfo  [0]{\@secondoftwo}%
\providecommand \bibfield  [0]{\@secondoftwo}%
\providecommand \translation [1]{[#1]}%
\providecommand \BibitemOpen [0]{}%
\providecommand \bibitemStop [0]{}%
\providecommand \bibitemNoStop [0]{.\EOS\space}%
\providecommand \EOS [0]{\spacefactor3000\relax}%
\providecommand \BibitemShut  [1]{\csname bibitem#1\endcsname}%
\let\auto@bib@innerbib\@empty
%</preamble>
\bibitem [{\citenamefont {Sachdev}\ \emph {et~al.}(2019)\citenamefont
  {Sachdev}, \citenamefont {Caudill}, \citenamefont {Fong}, \citenamefont {Lo},
  \citenamefont {Messick}, \citenamefont {Mukherjee}, \citenamefont {Magee},
  \citenamefont {Tsukada}, \citenamefont {Blackburn}, \citenamefont {Brady}
  \emph {et~al.}}]{sachdev2019gstlal}%
  \BibitemOpen
  \bibfield  {author} {\bibinfo {author} {\bibfnamefont {S.}~\bibnamefont
  {Sachdev}}, \bibinfo {author} {\bibfnamefont {S.}~\bibnamefont {Caudill}},
  \bibinfo {author} {\bibfnamefont {H.}~\bibnamefont {Fong}}, \bibinfo {author}
  {\bibfnamefont {R.~K.}\ \bibnamefont {Lo}}, \bibinfo {author} {\bibfnamefont
  {C.}~\bibnamefont {Messick}}, \bibinfo {author} {\bibfnamefont
  {D.}~\bibnamefont {Mukherjee}}, \bibinfo {author} {\bibfnamefont
  {R.}~\bibnamefont {Magee}}, \bibinfo {author} {\bibfnamefont
  {L.}~\bibnamefont {Tsukada}}, \bibinfo {author} {\bibfnamefont
  {K.}~\bibnamefont {Blackburn}}, \bibinfo {author} {\bibfnamefont
  {P.}~\bibnamefont {Brady}},  \emph {et~al.},\ }\href@noop {} {\bibfield
  {journal} {\bibinfo  {journal} {arXiv preprint arXiv:1901.08580}\ } (\bibinfo
  {year} {2019})}\BibitemShut {NoStop}%
\bibitem [{\citenamefont {Messick}\ \emph {et~al.}(2017)\citenamefont
  {Messick}, \citenamefont {Blackburn}, \citenamefont {Brady}, \citenamefont
  {Brockill}, \citenamefont {Cannon}, \citenamefont {Cariou}, \citenamefont
  {Caudill}, \citenamefont {Chamberlin}, \citenamefont {Creighton},
  \citenamefont {Everett} \emph {et~al.}}]{messick2017analysis}%
  \BibitemOpen
  \bibfield  {author} {\bibinfo {author} {\bibfnamefont {C.}~\bibnamefont
  {Messick}}, \bibinfo {author} {\bibfnamefont {K.}~\bibnamefont {Blackburn}},
  \bibinfo {author} {\bibfnamefont {P.}~\bibnamefont {Brady}}, \bibinfo
  {author} {\bibfnamefont {P.}~\bibnamefont {Brockill}}, \bibinfo {author}
  {\bibfnamefont {K.}~\bibnamefont {Cannon}}, \bibinfo {author} {\bibfnamefont
  {R.}~\bibnamefont {Cariou}}, \bibinfo {author} {\bibfnamefont
  {S.}~\bibnamefont {Caudill}}, \bibinfo {author} {\bibfnamefont {S.~J.}\
  \bibnamefont {Chamberlin}}, \bibinfo {author} {\bibfnamefont {J.~D.}\
  \bibnamefont {Creighton}}, \bibinfo {author} {\bibfnamefont {R.}~\bibnamefont
  {Everett}},  \emph {et~al.},\ }\href@noop {} {\bibfield  {journal} {\bibinfo
  {journal} {Physical Review D}\ }\textbf {\bibinfo {volume} {95}},\ \bibinfo
  {pages} {042001} (\bibinfo {year} {2017})}\BibitemShut {NoStop}%
\bibitem [{\citenamefont {Cannon}\ \emph {et~al.}(2012)\citenamefont {Cannon},
  \citenamefont {Cariou}, \citenamefont {Chapman}, \citenamefont
  {Crispin-Ortuzar}, \citenamefont {Fotopoulos}, \citenamefont {Frei},
  \citenamefont {Hanna}, \citenamefont {Kara}, \citenamefont {Keppel},
  \citenamefont {Liao} \emph {et~al.}}]{cannon2012toward}%
  \BibitemOpen
  \bibfield  {author} {\bibinfo {author} {\bibfnamefont {K.}~\bibnamefont
  {Cannon}}, \bibinfo {author} {\bibfnamefont {R.}~\bibnamefont {Cariou}},
  \bibinfo {author} {\bibfnamefont {A.}~\bibnamefont {Chapman}}, \bibinfo
  {author} {\bibfnamefont {M.}~\bibnamefont {Crispin-Ortuzar}}, \bibinfo
  {author} {\bibfnamefont {N.}~\bibnamefont {Fotopoulos}}, \bibinfo {author}
  {\bibfnamefont {M.}~\bibnamefont {Frei}}, \bibinfo {author} {\bibfnamefont
  {C.}~\bibnamefont {Hanna}}, \bibinfo {author} {\bibfnamefont
  {E.}~\bibnamefont {Kara}}, \bibinfo {author} {\bibfnamefont {D.}~\bibnamefont
  {Keppel}}, \bibinfo {author} {\bibfnamefont {L.}~\bibnamefont {Liao}},  \emph
  {et~al.},\ }\href@noop {} {\bibfield  {journal} {\bibinfo  {journal} {The
  Astrophysical Journal}\ }\textbf {\bibinfo {volume} {748}},\ \bibinfo {pages}
  {136} (\bibinfo {year} {2012})}\BibitemShut {NoStop}%
\bibitem [{\citenamefont {Team}(2020)}]{gstreamer}%
  \BibitemOpen
  \bibfield  {author} {\bibinfo {author} {\bibfnamefont {G.}~\bibnamefont
  {Team}},\ }\href@noop {} {\enquote {\bibinfo {title} {Gstreamer: open source
  multimedia framework},}\ } (\bibinfo {year} {2020})\BibitemShut {NoStop}%
\bibitem [{\citenamefont {Abbott}\ \emph {et~al.}(2017)\citenamefont {Abbott},
  \citenamefont {Abbott}, \citenamefont {Abbott}, \citenamefont {Acernese},
  \citenamefont {Ackley}, \citenamefont {Adams}, \citenamefont {Adams},
  \citenamefont {Addesso}, \citenamefont {Adhikari}, \citenamefont {Adya} \emph
  {et~al.}}]{abbott2017gw170817}%
  \BibitemOpen
  \bibfield  {author} {\bibinfo {author} {\bibfnamefont {B.~P.}\ \bibnamefont
  {Abbott}}, \bibinfo {author} {\bibfnamefont {R.}~\bibnamefont {Abbott}},
  \bibinfo {author} {\bibfnamefont {T.}~\bibnamefont {Abbott}}, \bibinfo
  {author} {\bibfnamefont {F.}~\bibnamefont {Acernese}}, \bibinfo {author}
  {\bibfnamefont {K.}~\bibnamefont {Ackley}}, \bibinfo {author} {\bibfnamefont
  {C.}~\bibnamefont {Adams}}, \bibinfo {author} {\bibfnamefont
  {T.}~\bibnamefont {Adams}}, \bibinfo {author} {\bibfnamefont
  {P.}~\bibnamefont {Addesso}}, \bibinfo {author} {\bibfnamefont
  {R.}~\bibnamefont {Adhikari}}, \bibinfo {author} {\bibfnamefont
  {V.}~\bibnamefont {Adya}},  \emph {et~al.},\ }\href@noop {} {\bibfield
  {journal} {\bibinfo  {journal} {Physical Review Letters}\ }\textbf {\bibinfo
  {volume} {119}},\ \bibinfo {pages} {161101} (\bibinfo {year}
  {2017})}\BibitemShut {NoStop}%
\bibitem [{\citenamefont {Abbott}\ \emph
  {et~al.}(2020{\natexlab{a}})\citenamefont {Abbott}, \citenamefont {Abbott},
  \citenamefont {Abbott}, \citenamefont {Abraham}, \citenamefont {Acernese},
  \citenamefont {Ackley}, \citenamefont {Adams}, \citenamefont {Adhikari},
  \citenamefont {Adya}, \citenamefont {Affeldt} \emph
  {et~al.}}]{abbott2020gw190425}%
  \BibitemOpen
  \bibfield  {author} {\bibinfo {author} {\bibfnamefont {B.}~\bibnamefont
  {Abbott}}, \bibinfo {author} {\bibfnamefont {R.}~\bibnamefont {Abbott}},
  \bibinfo {author} {\bibfnamefont {T.}~\bibnamefont {Abbott}}, \bibinfo
  {author} {\bibfnamefont {S.}~\bibnamefont {Abraham}}, \bibinfo {author}
  {\bibfnamefont {F.}~\bibnamefont {Acernese}}, \bibinfo {author}
  {\bibfnamefont {K.}~\bibnamefont {Ackley}}, \bibinfo {author} {\bibfnamefont
  {C.}~\bibnamefont {Adams}}, \bibinfo {author} {\bibfnamefont
  {R.}~\bibnamefont {Adhikari}}, \bibinfo {author} {\bibfnamefont
  {V.}~\bibnamefont {Adya}}, \bibinfo {author} {\bibfnamefont {C.}~\bibnamefont
  {Affeldt}},  \emph {et~al.},\ }\href@noop {} {\bibfield  {journal} {\bibinfo
  {journal} {The Astrophysical Journal Letters}\ }\textbf {\bibinfo {volume}
  {892}},\ \bibinfo {pages} {L3} (\bibinfo {year}
  {2020}{\natexlab{a}})}\BibitemShut {NoStop}%
\bibitem [{\citenamefont {Abbott}\ \emph {et~al.}(2016)\citenamefont {Abbott},
  \citenamefont {Abbott}, \citenamefont {Abbott}, \citenamefont {Abernathy},
  \citenamefont {Acernese}, \citenamefont {Ackley}, \citenamefont {Adams},
  \citenamefont {Adams}, \citenamefont {Addesso}, \citenamefont {Adhikari}
  \emph {et~al.}}]{abbott2016gw151226}%
  \BibitemOpen
  \bibfield  {author} {\bibinfo {author} {\bibfnamefont {B.~P.}\ \bibnamefont
  {Abbott}}, \bibinfo {author} {\bibfnamefont {R.}~\bibnamefont {Abbott}},
  \bibinfo {author} {\bibfnamefont {T.}~\bibnamefont {Abbott}}, \bibinfo
  {author} {\bibfnamefont {M.}~\bibnamefont {Abernathy}}, \bibinfo {author}
  {\bibfnamefont {F.}~\bibnamefont {Acernese}}, \bibinfo {author}
  {\bibfnamefont {K.}~\bibnamefont {Ackley}}, \bibinfo {author} {\bibfnamefont
  {C.}~\bibnamefont {Adams}}, \bibinfo {author} {\bibfnamefont
  {T.}~\bibnamefont {Adams}}, \bibinfo {author} {\bibfnamefont
  {P.}~\bibnamefont {Addesso}}, \bibinfo {author} {\bibfnamefont
  {R.}~\bibnamefont {Adhikari}},  \emph {et~al.},\ }\href@noop {} {\bibfield
  {journal} {\bibinfo  {journal} {Physical review letters}\ }\textbf {\bibinfo
  {volume} {116}},\ \bibinfo {pages} {241103} (\bibinfo {year}
  {2016})}\BibitemShut {NoStop}%
\bibitem [{\citenamefont {Nitz}\ \emph {et~al.}(2018)\citenamefont {Nitz},
  \citenamefont {Dal~Canton}, \citenamefont {Davis},\ and\ \citenamefont
  {Reyes}}]{nitz2018rapid}%
  \BibitemOpen
  \bibfield  {author} {\bibinfo {author} {\bibfnamefont {A.~H.}\ \bibnamefont
  {Nitz}}, \bibinfo {author} {\bibfnamefont {T.}~\bibnamefont {Dal~Canton}},
  \bibinfo {author} {\bibfnamefont {D.}~\bibnamefont {Davis}}, \ and\ \bibinfo
  {author} {\bibfnamefont {S.}~\bibnamefont {Reyes}},\ }\href@noop {}
  {\bibfield  {journal} {\bibinfo  {journal} {Physical Review D}\ }\textbf
  {\bibinfo {volume} {98}},\ \bibinfo {pages} {024050} (\bibinfo {year}
  {2018})}\BibitemShut {NoStop}%
\bibitem [{\citenamefont {Adams}\ \emph {et~al.}(2016)\citenamefont {Adams},
  \citenamefont {Buskulic}, \citenamefont {Germain}, \citenamefont {Guidi},
  \citenamefont {Marion}, \citenamefont {Montani}, \citenamefont {Mours},
  \citenamefont {Piergiovanni},\ and\ \citenamefont {Wang}}]{adams2016low}%
  \BibitemOpen
  \bibfield  {author} {\bibinfo {author} {\bibfnamefont {T.}~\bibnamefont
  {Adams}}, \bibinfo {author} {\bibfnamefont {D.}~\bibnamefont {Buskulic}},
  \bibinfo {author} {\bibfnamefont {V.}~\bibnamefont {Germain}}, \bibinfo
  {author} {\bibfnamefont {G.~M.}\ \bibnamefont {Guidi}}, \bibinfo {author}
  {\bibfnamefont {F.}~\bibnamefont {Marion}}, \bibinfo {author} {\bibfnamefont
  {M.}~\bibnamefont {Montani}}, \bibinfo {author} {\bibfnamefont
  {B.}~\bibnamefont {Mours}}, \bibinfo {author} {\bibfnamefont
  {F.}~\bibnamefont {Piergiovanni}}, \ and\ \bibinfo {author} {\bibfnamefont
  {G.}~\bibnamefont {Wang}},\ }\href@noop {} {\bibfield  {journal} {\bibinfo
  {journal} {Classical and Quantum Gravity}\ }\textbf {\bibinfo {volume}
  {33}},\ \bibinfo {pages} {175012} (\bibinfo {year} {2016})}\BibitemShut
  {NoStop}%
\bibitem [{\citenamefont {Hooper}\ \emph {et~al.}(2012)\citenamefont {Hooper},
  \citenamefont {Chung}, \citenamefont {Luan}, \citenamefont {Blair},
  \citenamefont {Chen},\ and\ \citenamefont {Wen}}]{hooper2012summed}%
  \BibitemOpen
  \bibfield  {author} {\bibinfo {author} {\bibfnamefont {S.}~\bibnamefont
  {Hooper}}, \bibinfo {author} {\bibfnamefont {S.~K.}\ \bibnamefont {Chung}},
  \bibinfo {author} {\bibfnamefont {J.}~\bibnamefont {Luan}}, \bibinfo {author}
  {\bibfnamefont {D.}~\bibnamefont {Blair}}, \bibinfo {author} {\bibfnamefont
  {Y.}~\bibnamefont {Chen}}, \ and\ \bibinfo {author} {\bibfnamefont
  {L.}~\bibnamefont {Wen}},\ }\href@noop {} {\bibfield  {journal} {\bibinfo
  {journal} {Physical Review D}\ }\textbf {\bibinfo {volume} {86}},\ \bibinfo
  {pages} {024012} (\bibinfo {year} {2012})}\BibitemShut {NoStop}%
\bibitem [{\citenamefont {Chu}(2017)}]{chu2017low}%
  \BibitemOpen
  \bibfield  {author} {\bibinfo {author} {\bibfnamefont {Q.}~\bibnamefont
  {Chu}},\ }\href@noop {} {\bibfield  {journal} {\bibinfo  {journal} {PhDT}\ }
  (\bibinfo {year} {2017})}\BibitemShut {NoStop}%
\bibitem [{\citenamefont {Aasi}\ \emph {et~al.}(2015)\citenamefont {Aasi},
  \citenamefont {Abbott}, \citenamefont {Abbott}, \citenamefont {Abbott},
  \citenamefont {Abernathy}, \citenamefont {Ackley}, \citenamefont {Adams},
  \citenamefont {Adams}, \citenamefont {Addesso}, \citenamefont {Adhikari}
  \emph {et~al.}}]{aasi2015advanced}%
  \BibitemOpen
  \bibfield  {author} {\bibinfo {author} {\bibfnamefont {J.}~\bibnamefont
  {Aasi}}, \bibinfo {author} {\bibfnamefont {B.}~\bibnamefont {Abbott}},
  \bibinfo {author} {\bibfnamefont {R.}~\bibnamefont {Abbott}}, \bibinfo
  {author} {\bibfnamefont {T.}~\bibnamefont {Abbott}}, \bibinfo {author}
  {\bibfnamefont {M.}~\bibnamefont {Abernathy}}, \bibinfo {author}
  {\bibfnamefont {K.}~\bibnamefont {Ackley}}, \bibinfo {author} {\bibfnamefont
  {C.}~\bibnamefont {Adams}}, \bibinfo {author} {\bibfnamefont
  {T.}~\bibnamefont {Adams}}, \bibinfo {author} {\bibfnamefont
  {P.}~\bibnamefont {Addesso}}, \bibinfo {author} {\bibfnamefont
  {R.}~\bibnamefont {Adhikari}},  \emph {et~al.},\ }\href@noop {} {\bibfield
  {journal} {\bibinfo  {journal} {Classical and quantum gravity}\ }\textbf
  {\bibinfo {volume} {32}},\ \bibinfo {pages} {074001} (\bibinfo {year}
  {2015})}\BibitemShut {NoStop}%
\bibitem [{\citenamefont {Acernese}\ \emph {et~al.}(2014)\citenamefont
  {Acernese}, \citenamefont {Agathos}, \citenamefont {Agatsuma}, \citenamefont
  {Aisa}, \citenamefont {Allemandou}, \citenamefont {Allocca}, \citenamefont
  {Amarni}, \citenamefont {Astone}, \citenamefont {Balestri}, \citenamefont
  {Ballardin} \emph {et~al.}}]{acernese2014advanced}%
  \BibitemOpen
  \bibfield  {author} {\bibinfo {author} {\bibfnamefont {F.}~\bibnamefont
  {Acernese}}, \bibinfo {author} {\bibfnamefont {M.}~\bibnamefont {Agathos}},
  \bibinfo {author} {\bibfnamefont {K.}~\bibnamefont {Agatsuma}}, \bibinfo
  {author} {\bibfnamefont {D.}~\bibnamefont {Aisa}}, \bibinfo {author}
  {\bibfnamefont {N.}~\bibnamefont {Allemandou}}, \bibinfo {author}
  {\bibfnamefont {A.}~\bibnamefont {Allocca}}, \bibinfo {author} {\bibfnamefont
  {J.}~\bibnamefont {Amarni}}, \bibinfo {author} {\bibfnamefont
  {P.}~\bibnamefont {Astone}}, \bibinfo {author} {\bibfnamefont
  {G.}~\bibnamefont {Balestri}}, \bibinfo {author} {\bibfnamefont
  {G.}~\bibnamefont {Ballardin}},  \emph {et~al.},\ }\href@noop {} {\bibfield
  {journal} {\bibinfo  {journal} {Classical and Quantum Gravity}\ }\textbf
  {\bibinfo {volume} {32}},\ \bibinfo {pages} {024001} (\bibinfo {year}
  {2014})}\BibitemShut {NoStop}%
\bibitem [{\citenamefont {{LIGO Scientific Collaboration, Virgo
  Collaboration}}(2019)}]{LIGOEMFOLLOWUSERGUIDE}%
  \BibitemOpen
  \bibfield  {author} {\bibinfo {author} {\bibnamefont {{LIGO Scientific
  Collaboration, Virgo Collaboration}}},\ }\href
  {https://emfollow.docs.ligo.org/userguide/content.html} {\bibfield  {journal}
  {\bibinfo  {journal} {Public Alerts User Guide}\ } (\bibinfo {year}
  {2019})}\BibitemShut {NoStop}%
\bibitem [{\citenamefont {Singer}\ and\ \citenamefont
  {Price}(2016)}]{singer2016rapid}%
  \BibitemOpen
  \bibfield  {author} {\bibinfo {author} {\bibfnamefont {L.~P.}\ \bibnamefont
  {Singer}}\ and\ \bibinfo {author} {\bibfnamefont {L.~R.}\ \bibnamefont
  {Price}},\ }\href@noop {} {\bibfield  {journal} {\bibinfo  {journal}
  {Physical Review D}\ }\textbf {\bibinfo {volume} {93}},\ \bibinfo {pages}
  {024013} (\bibinfo {year} {2016})}\BibitemShut {NoStop}%
\bibitem [{gra()}]{gracedb}%
  \BibitemOpen
  \href@noop {} {\enquote {\bibinfo {title} {Gravitational-wave candidate event
  database},}\ }\bibinfo {howpublished} {\url{https://gracedb.ligo.org}},\
  \bibinfo {note} {accessed: 2019-06-04}\BibitemShut {NoStop}%
\bibitem [{\citenamefont {Collaboration}\ \emph {et~al.}(2018)\citenamefont
  {Collaboration}, \citenamefont {Collaboration} \emph
  {et~al.}}]{ligo2018gwtc}%
  \BibitemOpen
  \bibfield  {author} {\bibinfo {author} {\bibfnamefont {L.~S.}\ \bibnamefont
  {Collaboration}}, \bibinfo {author} {\bibfnamefont {V.}~\bibnamefont
  {Collaboration}},  \emph {et~al.},\ }\href@noop {} {\bibfield  {journal}
  {\bibinfo  {journal} {arXiv preprint arXiv:1811.12907}\ } (\bibinfo {year}
  {2018})}\BibitemShut {NoStop}%
\bibitem [{\citenamefont {Abbott}\ \emph {et~al.}(2018)\citenamefont {Abbott}
  \emph {et~al.}}]{Aasi:2013wya}%
  \BibitemOpen
  \bibfield  {author} {\bibinfo {author} {\bibfnamefont {B.}~\bibnamefont
  {Abbott}} \emph {et~al.} (\bibinfo {collaboration} {KAGRA, LIGO Scientific,
  VIRGO}),\ }\href {\doibase 10.1007/s41114-018-0012-9} {\bibfield  {journal}
  {\bibinfo  {journal} {Living Rev. Rel.}\ }\textbf {\bibinfo {volume} {21}},\
  \bibinfo {pages} {3} (\bibinfo {year} {2018})},\ \Eprint
  {http://arxiv.org/abs/1304.0670} {arXiv:1304.0670 [gr-qc]} \BibitemShut
  {NoStop}%
\bibitem [{\citenamefont {Akutsu}\ \emph {et~al.}(2020)\citenamefont {Akutsu},
  \citenamefont {Ando}, \citenamefont {Arai}, \citenamefont {Arai},
  \citenamefont {Araki}, \citenamefont {Araya}, \citenamefont {Aritomi},
  \citenamefont {Aso}, \citenamefont {Bae}, \citenamefont {Bae} \emph
  {et~al.}}]{akutsu2020overview}%
  \BibitemOpen
  \bibfield  {author} {\bibinfo {author} {\bibfnamefont {T.}~\bibnamefont
  {Akutsu}}, \bibinfo {author} {\bibfnamefont {M.}~\bibnamefont {Ando}},
  \bibinfo {author} {\bibfnamefont {K.}~\bibnamefont {Arai}}, \bibinfo {author}
  {\bibfnamefont {Y.}~\bibnamefont {Arai}}, \bibinfo {author} {\bibfnamefont
  {S.}~\bibnamefont {Araki}}, \bibinfo {author} {\bibfnamefont
  {A.}~\bibnamefont {Araya}}, \bibinfo {author} {\bibfnamefont
  {N.}~\bibnamefont {Aritomi}}, \bibinfo {author} {\bibfnamefont
  {Y.}~\bibnamefont {Aso}}, \bibinfo {author} {\bibfnamefont {S.-W.}\
  \bibnamefont {Bae}}, \bibinfo {author} {\bibfnamefont {Y.-B.}\ \bibnamefont
  {Bae}},  \emph {et~al.},\ }\href@noop {} {\bibfield  {journal} {\bibinfo
  {journal} {arXiv preprint arXiv:2005.05574}\ } (\bibinfo {year}
  {2020})}\BibitemShut {NoStop}%
\bibitem [{\citenamefont {Cannon}\ \emph {et~al.}(2010)\citenamefont {Cannon},
  \citenamefont {Chapman}, \citenamefont {Hanna}, \citenamefont {Keppel},
  \citenamefont {Searle},\ and\ \citenamefont
  {Weinstein}}]{cannon2010singular}%
  \BibitemOpen
  \bibfield  {author} {\bibinfo {author} {\bibfnamefont {K.}~\bibnamefont
  {Cannon}}, \bibinfo {author} {\bibfnamefont {A.}~\bibnamefont {Chapman}},
  \bibinfo {author} {\bibfnamefont {C.}~\bibnamefont {Hanna}}, \bibinfo
  {author} {\bibfnamefont {D.}~\bibnamefont {Keppel}}, \bibinfo {author}
  {\bibfnamefont {A.~C.}\ \bibnamefont {Searle}}, \ and\ \bibinfo {author}
  {\bibfnamefont {A.~J.}\ \bibnamefont {Weinstein}},\ }\href@noop {} {\bibfield
   {journal} {\bibinfo  {journal} {Physical Review D}\ }\textbf {\bibinfo
  {volume} {82}},\ \bibinfo {pages} {044025} (\bibinfo {year}
  {2010})}\BibitemShut {NoStop}%
\bibitem [{pub()}]{publicnoise}%
  \BibitemOpen
  \href@noop {} {\enquote {\bibinfo {title} {Ligo document p1800374-v1},}\
  }\bibinfo {howpublished} {\url{https://dcc.ligo.org/P1800374/public/}},\
  \bibinfo {note} {accessed: 2020-10-26}\BibitemShut {NoStop}%
\bibitem [{\citenamefont {Collaboration}\ \emph {et~al.}(2020)\citenamefont
  {Collaboration}, \citenamefont {Collaboration} \emph
  {et~al.}}]{ligo2020gw190412}%
  \BibitemOpen
  \bibfield  {author} {\bibinfo {author} {\bibfnamefont {L.~S.}\ \bibnamefont
  {Collaboration}}, \bibinfo {author} {\bibfnamefont {V.}~\bibnamefont
  {Collaboration}},  \emph {et~al.},\ }\href@noop {} {\bibfield  {journal}
  {\bibinfo  {journal} {arXiv preprint arXiv:2004.08342}\ } (\bibinfo {year}
  {2020})}\BibitemShut {NoStop}%
\bibitem [{\citenamefont {Abbott}\ \emph
  {et~al.}(2020{\natexlab{b}})\citenamefont {Abbott}, \citenamefont {Abbott},
  \citenamefont {Abraham}, \citenamefont {Acernese}, \citenamefont {Ackley},
  \citenamefont {Adams}, \citenamefont {Adhikari}, \citenamefont {Adya},
  \citenamefont {Affeldt}, \citenamefont {Agathos} \emph
  {et~al.}}]{abbott2020gw190521}%
  \BibitemOpen
  \bibfield  {author} {\bibinfo {author} {\bibfnamefont {R.}~\bibnamefont
  {Abbott}}, \bibinfo {author} {\bibfnamefont {T.}~\bibnamefont {Abbott}},
  \bibinfo {author} {\bibfnamefont {S.}~\bibnamefont {Abraham}}, \bibinfo
  {author} {\bibfnamefont {F.}~\bibnamefont {Acernese}}, \bibinfo {author}
  {\bibfnamefont {K.}~\bibnamefont {Ackley}}, \bibinfo {author} {\bibfnamefont
  {C.}~\bibnamefont {Adams}}, \bibinfo {author} {\bibfnamefont
  {R.}~\bibnamefont {Adhikari}}, \bibinfo {author} {\bibfnamefont
  {V.}~\bibnamefont {Adya}}, \bibinfo {author} {\bibfnamefont {C.}~\bibnamefont
  {Affeldt}}, \bibinfo {author} {\bibfnamefont {M.}~\bibnamefont {Agathos}},
  \emph {et~al.},\ }\href@noop {} {\bibfield  {journal} {\bibinfo  {journal}
  {Physical review letters}\ }\textbf {\bibinfo {volume} {125}},\ \bibinfo
  {pages} {101102} (\bibinfo {year} {2020}{\natexlab{b}})}\BibitemShut
  {NoStop}%
\bibitem [{\citenamefont {Abbott}\ \emph
  {et~al.}(2020{\natexlab{c}})\citenamefont {Abbott}, \citenamefont {Abbott},
  \citenamefont {Abraham}, \citenamefont {Acernese}, \citenamefont {Ackley},
  \citenamefont {Adams}, \citenamefont {Adhikari}, \citenamefont {Adya},
  \citenamefont {Affeldt}, \citenamefont {Agathos} \emph
  {et~al.}}]{abbott2020gw190814}%
  \BibitemOpen
  \bibfield  {author} {\bibinfo {author} {\bibfnamefont {R.}~\bibnamefont
  {Abbott}}, \bibinfo {author} {\bibfnamefont {T.}~\bibnamefont {Abbott}},
  \bibinfo {author} {\bibfnamefont {S.}~\bibnamefont {Abraham}}, \bibinfo
  {author} {\bibfnamefont {F.}~\bibnamefont {Acernese}}, \bibinfo {author}
  {\bibfnamefont {K.}~\bibnamefont {Ackley}}, \bibinfo {author} {\bibfnamefont
  {C.}~\bibnamefont {Adams}}, \bibinfo {author} {\bibfnamefont
  {R.}~\bibnamefont {Adhikari}}, \bibinfo {author} {\bibfnamefont
  {V.}~\bibnamefont {Adya}}, \bibinfo {author} {\bibfnamefont {C.}~\bibnamefont
  {Affeldt}}, \bibinfo {author} {\bibfnamefont {M.}~\bibnamefont {Agathos}},
  \emph {et~al.},\ }\href@noop {} {\bibfield  {journal} {\bibinfo  {journal}
  {The Astrophysical Journal Letters}\ }\textbf {\bibinfo {volume} {896}},\
  \bibinfo {pages} {L44} (\bibinfo {year} {2020}{\natexlab{c}})}\BibitemShut
  {NoStop}%
\end{thebibliography}%

\end{document}